# Monitoring of Particle Flux and LET Variations with Pulse Stretching Inverters

Marko Andjelkovic, Junchao Chen, Aleksandar Simevski, Oliver Schrape, Milos Krstic, Rolf Kraemer

*Abstract*—**This work investigates the use of pulse stretching inverters for monitoring the variation of flux and Linear Energy Transfer (LET) of energetic particles. The basic particle detector consists of two cascaded pulse stretching (skew-sized) inverters designed in CMOS technology, and the required sensing area is obtained by connecting multiple two-inverter pulse stretching cells in parallel, and employing the required number of parallel arrays. The particle strikes are detected in terms of the Single Event Transients (SETs), and the detector provides the information on the SET count rate and SET pulse width variation, from which the particle flux and LET can be determined. The main advantage of the proposed solution is the possibility to sense the LET variations using purely digital processing logic. The SPICE simulations done on IHP's 130 nm bulk CMOS technology have shown that the SET pulse width at the output of detector changes by 550 ps over the LET range from 1 to 100 MeVcm$^2$mg$^{-1}$. The proposed solution is intended to operate as an on-chip particle detector within the self-adaptive multiprocessing systems.**

## I. INTRODUCTION

THE integrated circuits (ICs) employed in space missions are exposed to intense bursts of energetic particles which can causes soft errors and possibly result in failure of the entire system. As the space missions are usually long-term, the radiation hardness of space-borne electronics is one of the main design goals. Because the radiation intensity in space can vary over several orders of magnitude and the peak exposure levels can last for hours or days [1], the electronic systems require the dynamic hardening mechanisms which can be activated only under the critical radiation levels. In that way, the performance and power consumption penalties can be minimized and the system resources can be efficiently utilized according to the application requirements. A practical implementation of such concept is a self-adaptive multi-core processing system, where the cores are dynamically configured into different operating modes such as high performance mode (all cores are executing different tasks), de-stress/low-power mode (some cores are switched off) and fault-tolerant mode (cores are arranged as N-modular redundant subsystems) [2].

To enable the dynamic activation of radiation hardening mechanisms within an electronic system employed in space, the real-time monitoring of flux and Linear Energy Transfer (LET) of energetic particles, such as heavy ions, is mandatory. The system soft error rate (SER) increases linearly with the particle flux. Moreover, the high-LET particles can cause multi-bit upsets in sequential and memory elements, as well as long single event transients (SETs) in combinational logic [3], which increases the probability of soft error occurrence. In particular, with the CMOS technology scaling, increase of operating frequency and reduction of supply voltage and logic depth, the SET contribution to the system SER increases [4]. Therefore, neglecting the impact of particles' LET on the system reliability may severely underestimate the SER. By detecting the real-time variation of particle flux and LET, the corresponding variation of the system SER can be estimated during the run-time, and the appropriate hardening mechanisms can be activated to reduce the SER and ensure the reliable system operation.

The widely used particle detectors, providing very good accuracy and resolution in measuring the flux and LET, are the diode-based detectors [5]. However, the need for analog processing increases the complexity and cost, particularly if the detector has to be integrated on the chip. A cost-effective alternative are the SRAM-based detectors [6 – 8] which can measure the flux in terms on the number of upsets, but cannot detect the LET. A custom-designed SRAM with all cells connected in parallel has been used for measuring the flux and LET [9]. However, this solution also requires analog processing since the LET is determined from the amplitude of induced SETs. The alternative detectors such as bulk-built current detectors [10, 11] and acoustic wave detectors [12] have been proposed, but neither of them supports the LET measurement. Recently, the use of 3D NAND flash memory as a particle detector has been reported [13]. In this case, the LET is determined from the threshold voltage shift of floating gate transistors, which also requires analog processing. Moreover, the 3D NAND flash memories are still not fully adopted in space applications. All aforementioned solutions have been implemented as stand-alone monitors. To the best of our knowledge, only two works [1, 14] have reported the use of embedded memories as particle detectors for self-adaptive systems for space applications, but in both cases only the flux measurement is supported.

In this work, we analyze an alternative approach for monitor-

M. Andjelkovic, J. Chen, A. Simevski, O. Schrape, M. Krstic and R. Kraemer are with IHP – Leibniz Institut für innovative Mikroelekronik, Frankfurt Oder, Germany. M. Krstic is also with University of Potsdam, Potsdam, Germany. R. Kraemer is also with Brandenburg University of Technology, Cottbus, Germany. Emails: {andjelkovic, chen, simevski, schrape, krstic, kraemer}@ihp-microelectronics.com.

ing of particle flux and LET variations, with the benefits of simple processing logic and the possibility of integration into the target IC. Our approach is based on the use of custom-sized pulse stretching (skewed) inverters for detecting the incident energetic particles in terms of induced SETs. The use of pulse stretching inverters for measuring the particle flux has been initially proposed in our previous work [15]. Here, we introduce an enhancement of initial design to support the detection of both flux and LET variations.

The rest of the paper is organized as follows. Section II gives a brief description of the concept of particle detection with the pulse stretching inverters and introduces the parallel configuration of the pulse stretching cells. The simulation analysis of the SET effects in the proposed detector is given in Section III. The general design of the readout circuit is presented in Section IV, and an application in a self-adaptive multiprocessing system is presented in Section V.

## II. Pulse Stretching Inverters as Particle Detectors

The skewed logic has been traditionally used in high speed circuits due to the fast signal transitions. A specific benefit of skewed logic that can be exploited for detection of energetic particles is the pulse stretching feature [15]. For that reason, we use here the expression “pulse stretching logic” instead of “skewed logic”. The simplest pulse stretching circuit in CMOS technology can be realized with two inverters as depicted in Figure 1. In the following discussion, the two-inverter pulse stretcher will be denoted as the Pulse Stretching Cell (PSC).

By applying a fixed low level at the input of a PSC, NMOS1 and PMOS2 will be sensitive to particle strikes, while PMOS1 and NMOS2 will act as restoring elements. The sensitive (off-state) transistors are sized by increasing the channel width, to increase the sensitive area. The restoring (on-state) transistors are sized by decreasing the channel width and increasing the channel length, to increase the sensitivity to a single particle strike. As shown in [15], with appropriate transistor sizing the sensitivity of a PSC can be adjusted such that the low-LET particles can induce the SETs in the order of ns.

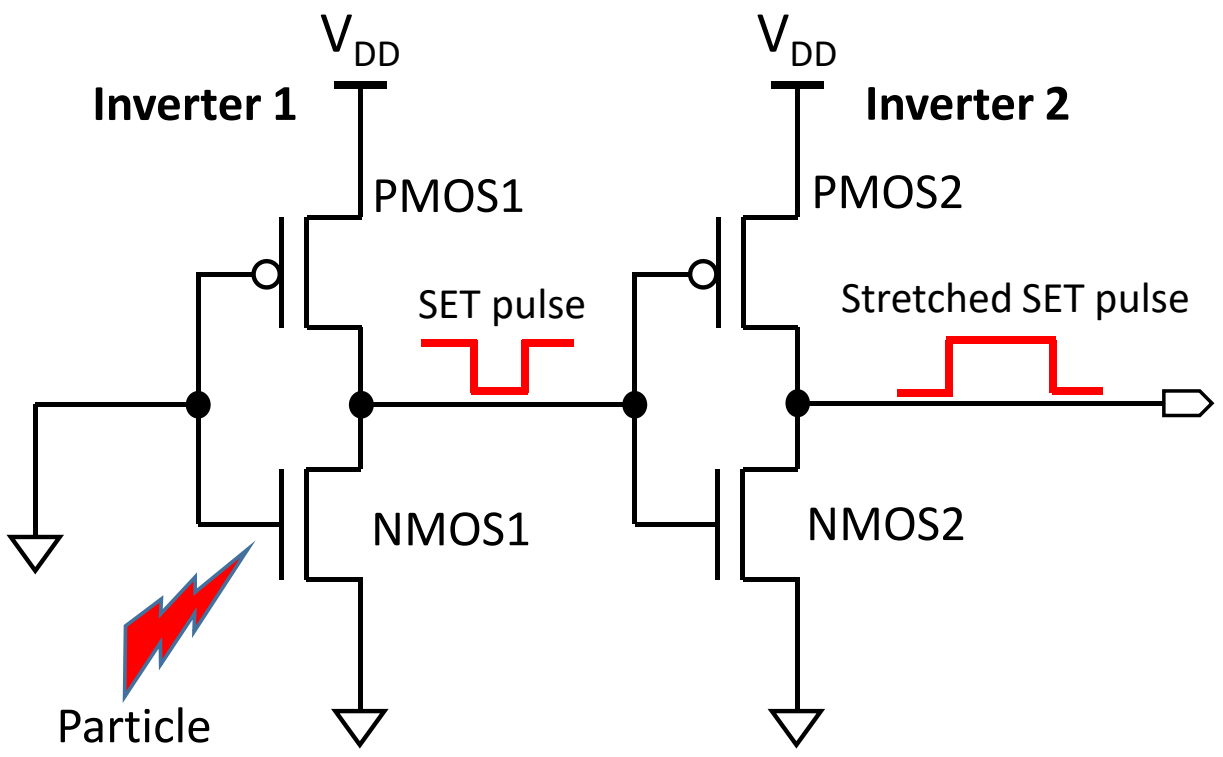

Figure 1: Two-inverter pulse stretching cell (PSC)

Although long chains of serially connected PSCs [15] are suitable for capturing the SET count rate, from which the particle flux can be determined, this approach cannot detect the LET variations for two reasons. First, in a long pulse stretching chain the SET pulse width saturates at very low LETs and is independent of LET variation over a wide range. Second, in a long chain, the two particles with different LETs can cause similar SET pulse widths if different PSC stages are hit, thus making it difficult or even impossible to differentiate the circuit response due to these two strikes. An additional drawback of serially connected PSC chain is high latency since the induced SET pulse can be stretched beyond 40 ns.

To resolve the aforementioned issues of long PSC chains, we propose a parallel connection of multiple PSCs for detecting both the flux and LET variation. However, the number of PSC cells that can be connected in parallel is limited due to the load effect, i.e. by increasing the number of connected cells, the load for each cell is increased and this limits the sensitivity to particle strikes. Therefore, to further increase the sensitive area, multiple arrays can be used, as shown in Figure 2, and the arrays are then connected with OR-tree.

The measurement of particle flux with the proposed detector is similar to other detectors. Basically, the number of detected SETs ($N_{SET}$) in a time period $T$ is proportional to flux $\Phi$. With the cross-section $CS$ of the detector obtained from radiation experiment, the flux can be calculated as $\Phi = N_{SET} / (CS \times T)$. As $CS$ varies with LET, the saturation value of $CS$ can be used to determine the flux. On the other hand, the proposed detector does not support the measuring of actual LET, but it rather enables to detect the LET variations in terms of the SET pulse width variation. The measured SET pulse widths are sorted into several distinct ranges which can be used as a representation of actual LET ranges.

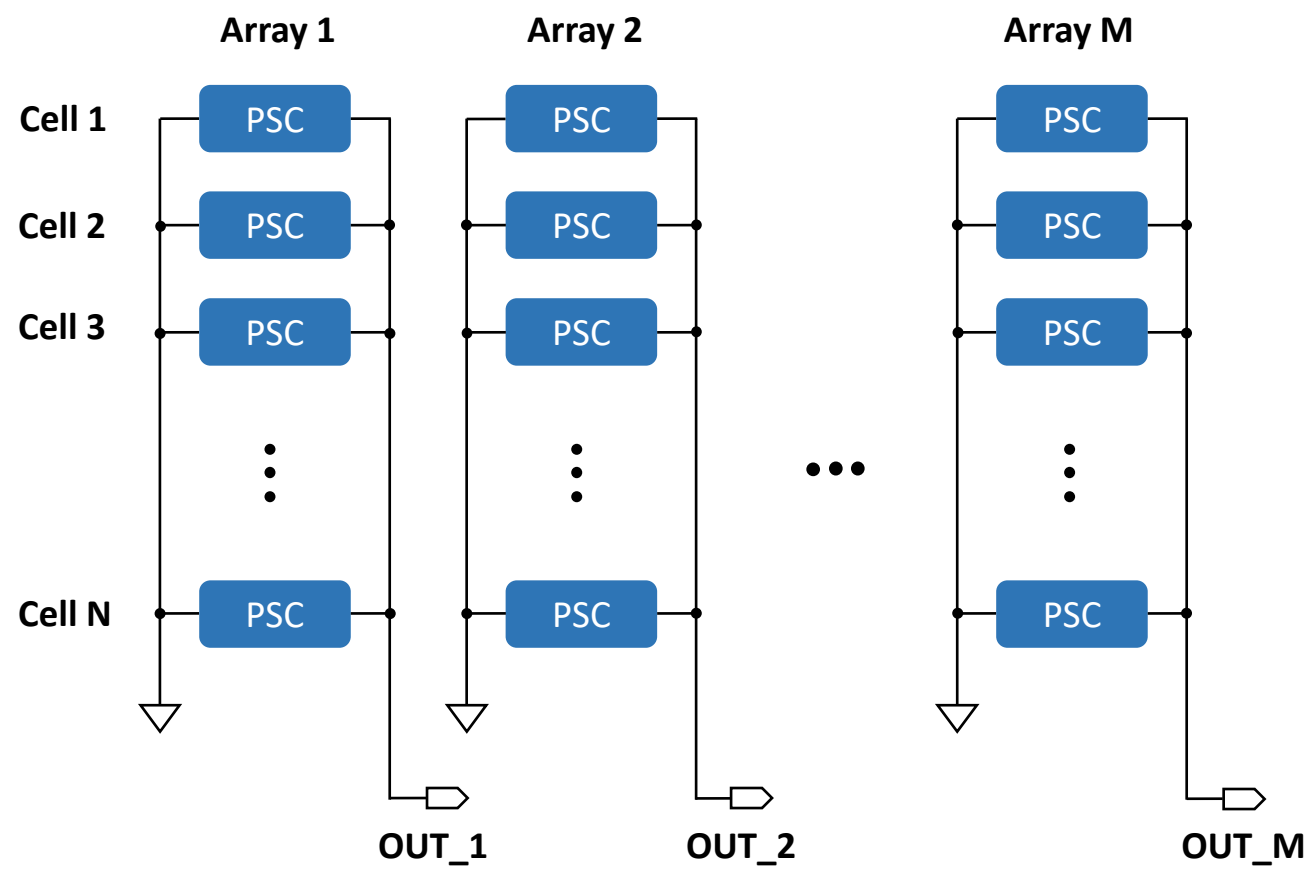

Figure 2: Proposed particle detector based on multiple arrays of pulse stretching cells (PSC) connected in parallel

## III. Simulation of SET Effects in PSC Arrays

The circuit simulations were done for IHP’s 130 nm technology, with Cadence Spectre. This technology is robust to single event latchup (SEL) up to 67 MeVcm$^2$mg$^{-1}$ and to total ionizing dose (TID) up to 900 krad [16]. A bias-dependent current model [17], implemented as Verilog-A model, was used for simulation of SETs. The LET was varied during simulations, for the fixed rise and fall time constants of 10 ps and 100 ps, respectively. In comparison to double-exponential model, the bias-dependent model takes into account the impact of node bias on the SET current and hence reproduces more accurately the SET pulse.

Since the proposed detector consists of a number of identical arrays, it is sufficient to perform the analysis only for one array. Initially, the exhaustive circuit simulations were done to determine the transistors sizes for a single PSC, in order to obtain as high sensitivity as possible. In Table 1, the chosen channel widths and lengths for both transistors in the PSC are given. These transistor sizes enable to detect the SET pulses caused by particles with LET below 1 $MeVcm^2mg^{-1}$. Using the selected transistor sizes, the SET pulse width at the output of a PSC array was analyzed in terms of: (i) number of PSCs in parallel, (ii) particle LET, (iii) supply voltage, and (iv) temperature.

Table 1: Transistor sizes for a pulse stretching cell (PSC)

| Inverter in PSC | Channel width and length |
|---|---|
| Inverter 1 | $W_{PMOS1}$ = 150 nm, $L_{PMOS1}$ = 1 µm<br>$W_{NMOS1}$ = 6 µm, $L_{NMOS1}$ = 130 nm |
| Inverter 2 | $W_{PMOS2}$ = 6 µm, $L_{PMOS2}$ = 130 nm<br>$W_{NMOS2}$ = 150 nm, $L_{NMOS2}$ = 6 µm |

The SET pulse width at the output of PSC, as a function of LET and the number of PSCs in connected parallel, is illustrated in Figure 3 (for SETs induced in Inverter 1) and Figure 4 (for SETs induced in Inverter 2). The presented results have been obtained for nominal supply voltage of 1.2 V and temperature of 27 ºC. The output SET pulse width increases logarithmically with LET (the dependence appears as linear due to log-scale on x-axis). By increasing the number of cells in parallel, the load capacitance for each cell increases and thus the output SET pulse width decreases. For both inverters, the SET pulse width increases by almost 550 ps in the LET range from 1 to 100 $MeVcm^2mg^{-1}$. This LET range covers most of the real particles that can be encountered in space. The output SET is wider when Inverter 1 is hit because the induced SET pulse is stretched while propagating through Inverter 2. For more than 12 PSCs in parallel, the SETs in Inverter 2 resulting from low-LET particles cannot be detected. As can be seen in Figure 4, for 12 PSCs connected in parallel the SET pulse width due to strike in Inverter 2 decreases significantly for LET < 5 $MeVcm^2mg^{-1}$. Therefore, for investigated design, the maximum number of PSCs that can be connected in parallel is 12.

Based on the results depicted in Figures 3 and 4, the SET pulse widths for a 12-cell array can be classified in three ranges, as shown in Table 2. The SET ranges correspond to the investigated LET ranges. Due to the large gap between the SET pulse width ranges for the two inverters of a PSC, the risk of ambiguous interpretation of the SET pulse width is minimized. However, as the SET pulse width depends on the particle strike location, the distribution of SET pulse widths will be obtained for each LET. As demonstrated in [18], for different particle strike locations the SET pulse width in 130 nm technology can vary by more than 400 ps. Hence, only with the TCAD or experimental calibration is possible to define the authentic SET pulse width range for a particular LET range. Nevertheless, the proposed concept of SET width range classification is flexible and can be adopted to different SET widths.

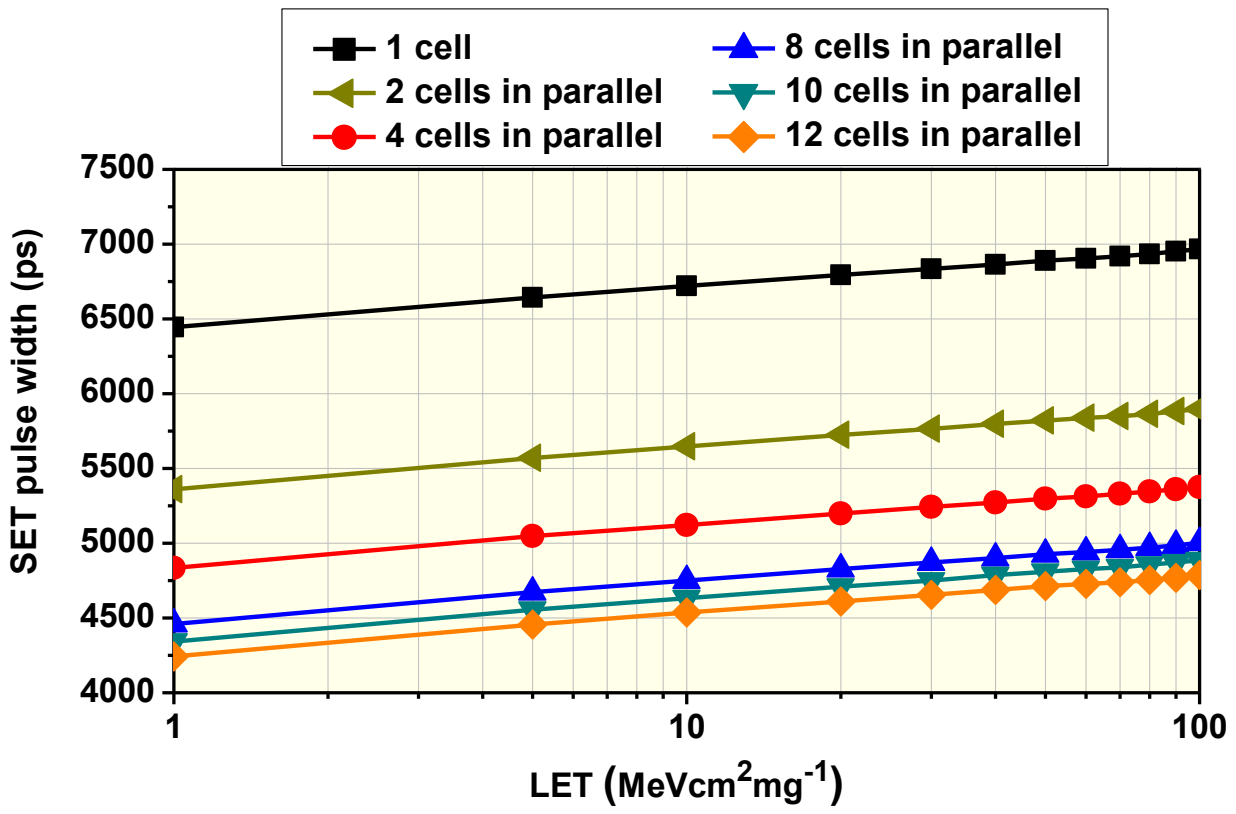

Figure 3: SET pulse width as a function of LET and number of cells connected in parallel, resulting from current injection in Inverter 1 of PSC

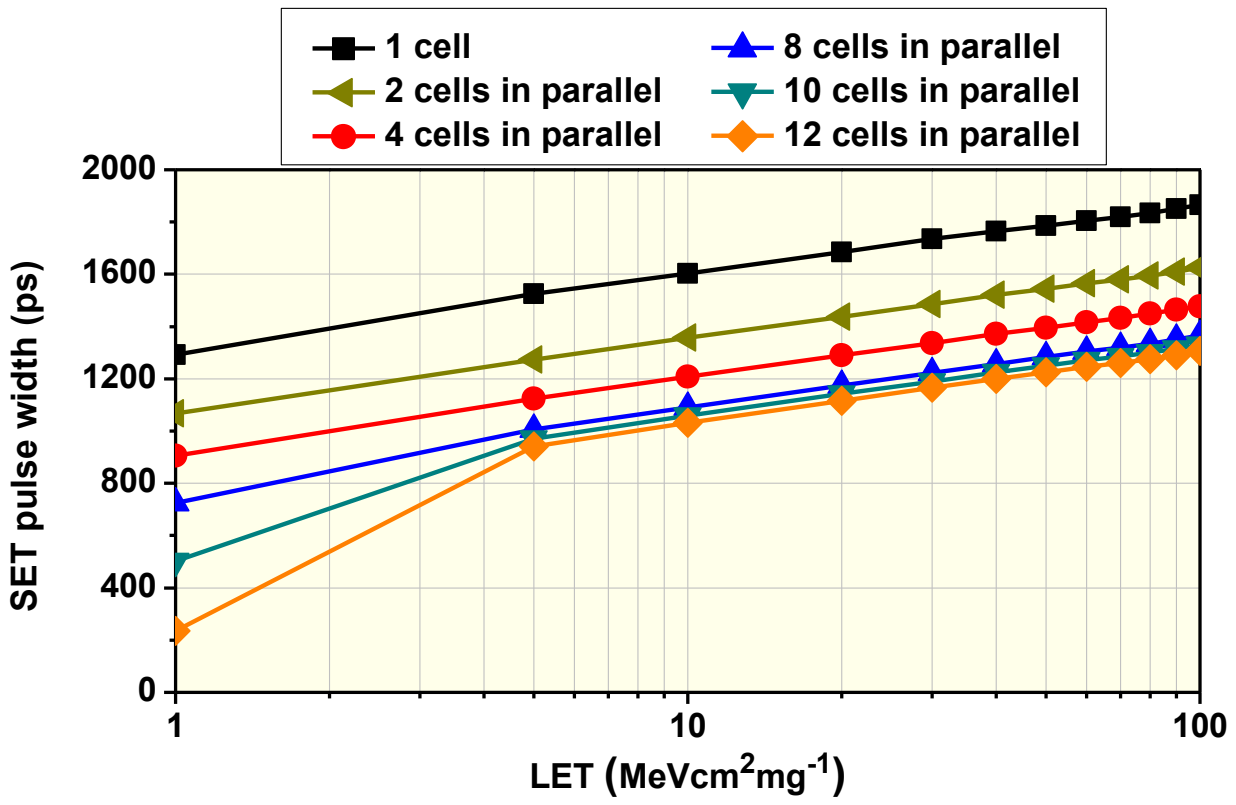

Figure 4: SET pulse width as a function of LET and number of cells connected in parallel, resulting from current injection in Inverter 2 of PSC

Table 2: SET pulse width ranges for a PSC in a 12-cell array

| Criticality level | LET range ($MeVcm^2mg^{-1}$) | SET pulse width range (ns) | |
|---|---|---|---|
| | | Inverter 1 | Inverter 2 |
| Low | 1 – 5 | 4.24 – 4.46 | 0.24 - 0.94 |
| Medium | 5 – 30 | 4.46 – 4.65 | 0.94 – 1.17 |
| High | 30 – 100 | 4.65 – 4.79 | 1.17 – 1.30 |

Apart from LET and the number of PSCs connected in parallel, the SET pulse width is also influenced by the supply voltage and temperature. At lower supply voltage, the threshold LET is reduced and the induced SET pulses are wider. The comparison of the change of SET pulse width at the output of a 12-cell array, when the current is successively injected in the two transistors of a PSC, is shown in Figure 5. The illustrated results are for LET = 10 $MeVcm^2mg^{-1}$. As can be seen, at supply voltage of 0.8 V the SET pulse width due to strikes in Inverter 1 is wider by almost 2.8 ns compared to that for nominal supply voltage, and by almost 800 ps for strikes in Inverter 2. Thus, the voltage scaling can be utilized as a means for controlling the sensitivity of PSC to particle strikes.

The impact of temperature on the SET pulse width at the output of a PSC in a 12-cell array, for LET = 10 MeVcm$^2$mg$^{-1}$, is illustrated in Figure 6. The results illustrate the change of SET pulse width as a function of the change of temperature with respect to 27 $^0$C. As can be seen, lower temperatures result in decrease of the SET pulse width, while at higher temperatures the SET pulse width increases. It is particularly important to note that the temperature variation have significantly stronger impact for SETs induced in Inverter 1, as the SET pulse width changes by almost 1.5 ns.

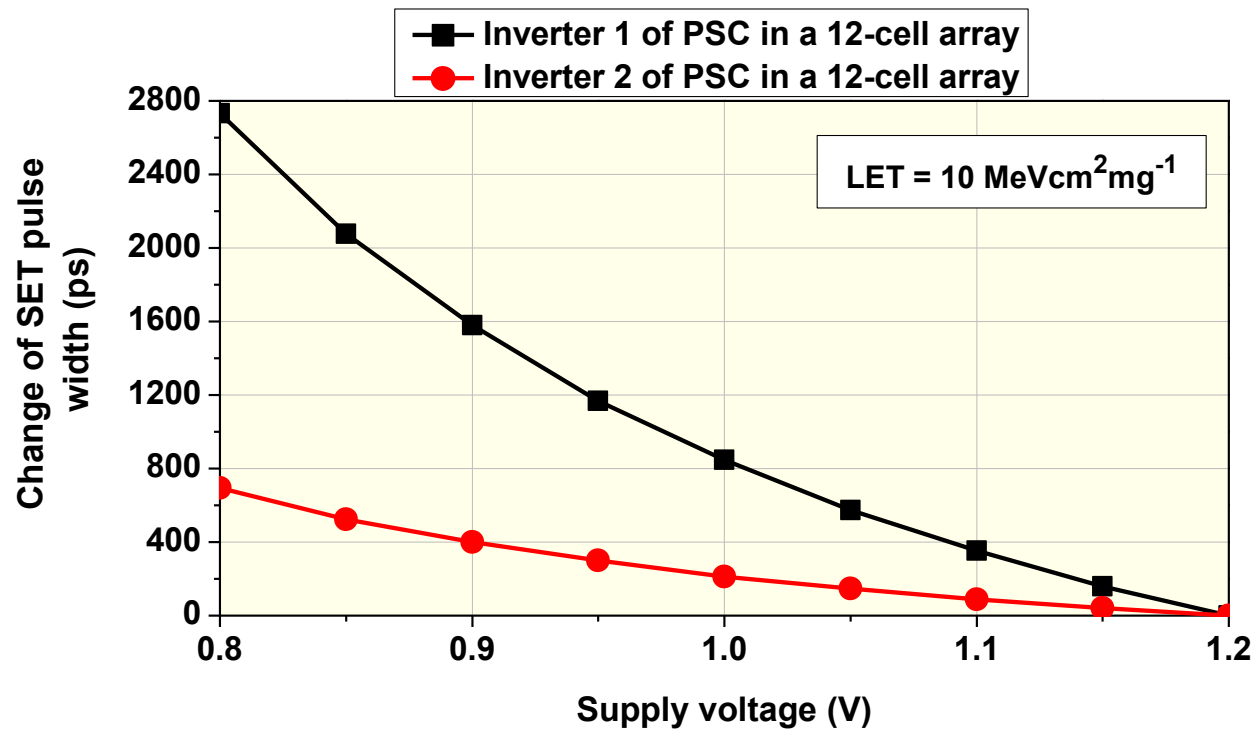

Figure 5: Change of SET pulse width at the output of a 12-cell parallel array as a function of supply voltage, for LET = 10 MeVcm$^2$mg$^{-1}$

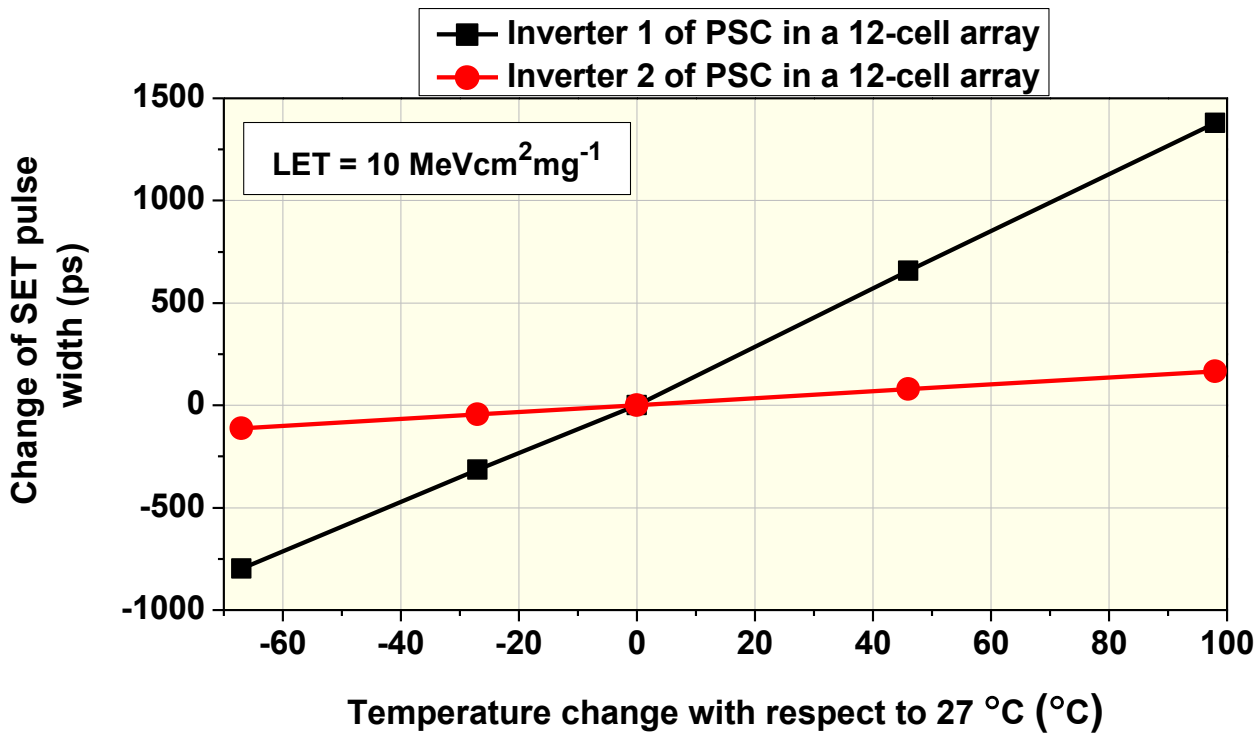

Figure 6: Change of SET pulse width at the output of a 12-cell parallel array as a function of temperature change, for LET = 10 MeVcm$^2$mg$^{-1}$

## IV. Readout Circuit

In comparison to diode and 3D NAND flash detectors which require analog logic for measuring the particle flux and LET, our solution employs a simple purely digital readout circuit as illustrated in Figure 7. The circuit performs two tasks: counting of detected SETs and sorting the detected SETs into pulse width ranges. An important feature is the low power consumption, since the logic level switching in the detector, OR-tree, filters and counters occurs only when an SET is generated.

The outputs of all pulse stretching arrays are fed to an OR-tree to obtain a single output signal for further processing. The output of the OR-tree is interfaced to the SET filters, where each filter passes the SET pulses within a specific width range. Each filter is connected to a ripple counter which captures the number of detected SETs in the corresponding SET width range. The filters are based on standard delay logic. The size of the OR-tree is chosen according to the number of arrays, such that the number of OR gates is less than 10 % of the number of PSC cells. On the other hand, the number of PSC cells and arrays depends on the required sensing area. This is the benefit of the proposed approach, since the detector scalability allows for easier integration into a target chip.

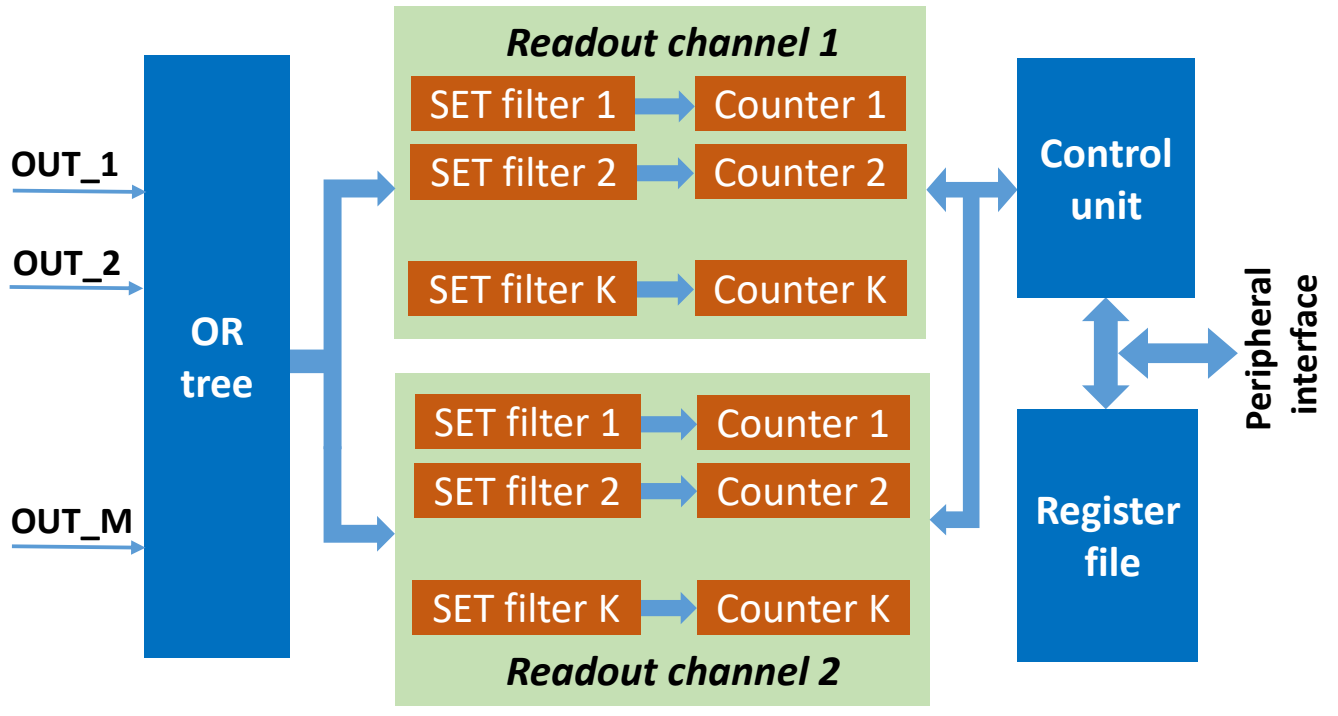

Figure 7: Block diagram of readout circuit for pulse stretching detector

As the SETs induced in a single PSC can be sorted in two distinct pulse width ranges (as shown in Table 2), depending on whether the first or second inverter is hit, the filters and counters are divided into two channels (*Readout channel 1* and *Readout channel 2*), as shown in Figure 7. Each channel is composed of $K$ filters and $K$ counters, where $K$ denotes the number of SET width ranges ($K$ = 3 for the classification as in Table 2). The counter states of the equivalent ranges from the two channels are summed. Thus, for a predefined measurement period $T$, three values will be acquired by the processing unit, where each value represents the number of detected SETs ($N_{SET}$) for one of the three SET width ranges.

## V. Application in a Self-Adaptive System

Since the target application of the proposed particle monitor are the self-adaptive multi-core systems for space missions, it is essential to establish a correlation between the response of the response and the SER of the target system. Although the detector cannot measure the actual LET, the information on the SET count rate and the classification of SETs into pulses width ranges is sufficient for estimating the criticality of the radiation exposure for a target system. With this information, the operating modes and the fault-tolerance mechanisms can be adjusted during the runtime.

In principle, for any multi-core system, the operating modes of each core are changing during the runtime and this leads to the change of the overall SER. As the SER of a system is the sum of the SER contribution of all cores, the SER of a single core composed of $N$ standard logic units can be expressed as,

$$SER_{CORE} = \Phi \times \sum_{i=1}^{N} SER_{GEN}(i) \times SER_{DER}(i) \qquad (1)$$

where $\Phi$ is the particle flux, $SER_{GEN}$ is the soft error generation probability and $SER_{DER}$ is the product of derating factors such as electrical, logical and timing derating. The design and technology parameters are constant for a given system. Thus, the overall SER for a particular operating mode (task) will depend on irradiation parameters (flux and LET), operating parameters (supply voltage, frequency, temperature and logic states of combinational and sequential gates) and sensitive area (i.e. total area of all off-state transistors).

The impact of LET is particularly important for combinational logic, as it defines the SET pulse width which is linearly related to the timing masking probability. However, due to the stretching feature of the detector, the measured SET pulse widths are not the SET widths that would be induced in standard cells. The SET pulse widths in standard cells in 130 nm technology are typically in the order of hundreds of ps, as shown in [17]. If the standard cells are characterized with the same simulation setup, the SET widths for each cell can be mapped to equivalent SET widths captured by the detector. Table 3 shows the simulated SET pulse widths for the three common logic gates (for low input levels), corresponding to the criticality levels defined in Table 2.

Table 3: SET width ranges for standard cells in 130 nm technology

| **Criticality level** | **SET pulse width for standard logic cells (ns)** | | |
|---|---|---|---|
| | **INV** | **NAND** | **NOR** |
| Low | < 0.30 | < 0.28 | < 0.24 |
| Medium | 0.30 – 0.49 | 0.28 – 0.47 | 0.24 – 0.44 |
| High | > 0.49 | > 0.47 | > 0.44 |

According to the operating conditions and detected flux and LET, the SER contribution of the most critical combinational and sequential cells in the system can be assessed in real time. The critical (most sensitive) cells have to be identified in the design phase. Then, a variety of system-level hardening measures, such as voltage and/or frequency scaling, Dual Modular Redundancy (DMR) and Triple Modular Redundancy (TMR), can be activated dynamically during the runtime. To enable this functionality, along with the particle monitoring is necessary to implement a custom-designed processing unit for calculation of SER in real-time.

## VI. Conclusion

In this paper, a unique design of a particle detector based on pulse stretching cells, for the real-time monitoring of particle flux and LET variations, is presented. The proposed solution is intended to serve as an enabler of the dynamic fault tolerance mechanisms in a self-adaptive multiprocessor system. The SPICE simulations based on the current injection with a bias-dependent current model have shown that for LET from 1 to 100 $MeVcm^2mg^{-1}$, the SET pulse width varies by approximately 550 ps. The LET variations can be classified into three ranges, where each range corresponds to a unique SET pulse width range. A simple and low power digital readout logic can be integrated with the detector in a target chip, thus allowing to sense the exact radiation conditions to which the target system is exposed. The main objectives of future work will perform irradiation experiments in order to calibrate the detector, and establish an approach for the real-time assessment of the target system's SER variation.

## Acknowledgment

This work was done in the framework of project REDOX (KR 3576/29-1), funded by the German Research Foundation DFG (*Deutsche Forschungsgemeinschaft*).

## References

[1] R. Glein et al., "Detection of Solar Particle Events inside FPGAs," in Proc. European Conference on Radiation Effects on Components and Systems (RADECS), 2016.
[2] A. Simevski, et al., "PISA: Power-robust Microprocessor Design for Space Applications," in Proc. IEEE International Symposium on Online Testing and Robust System Design (IOLTS), 2020.
[3] P. E. Dodd et al., "Production and Propagation of Single Event Transients in High-Speed Digital Logic ICs," IEEE Trans on Nuclear Science, 2005.
[4] N. N. Mahatme et al., "Impact of Technology Scaling on the Combinational Logic Soft Error Rate," in Proc. IEEE International Reliability Physics Symposium (IRPS), 2014.
[5] W. S. Wong et al, "Introducing Timepix2, A Frame-Based Pixel Detector Readout ASIC Measuring Energy Deposition and Arrival Time," Radiation Measurements, 2020.
[6] R. Harboe-Sorensen et al., "Design, Testing and Calibration of a Reference SEU Monitor System," in Proc. European Conference on Radiation Effects on Components and Systems (RADECS), 2005.
[7] G. Tsiligiannis et al., "An SRAM Based Monitor for Mixed-Field Radiation Environments," IEEE Trans. on Nuclear Science, 2014.
[8] J, Prinzie et al., "An SRAM-Based Radiation Monitor with Dynamic Voltage Control in 0.18 µm CMOS Technology," IEEE Trans. on Nuclear Science, 2019.
[9] E. G. Stassinopoulos et al., "Measurement of Cosmic Ray and Trapped Proton LET Spectra on the STS-95 HOST Mission," IEEE Trans. on Nuclear Science, 2017.
[10] E. H. Neto, I. Ribeiro, M. Vieira, G. Wirth, F. L. Kastensmidt, "Evaluating Fault Coverage of Bulk Built-in Current Sensor for Soft Errors in Combinational and Sequential Logic," in Proc. Symposium on Integrated Circuits and Systems Design (SBCCI), 2005.
[11] R. Possamai Basstos et al., "Assessment of On-Chip Current Sensor for Detection of Thermal-Neutron Induced Transients," IEEE Trans. on Nuclear Science, 2020.
[12] G. Upasani, H. Vera, A. Gonzales, "A Case for Acoustic Wave Detectors for Soft-Errors," IEEE Trans. on Computers, 2018.
[13] M. Bagatin et al., "A Heavy-Ion Detector Based on 3-D NAND Flash Memories," IEEE Trans. on Nuclear Science, 2020.
[14] J. Chen et al., "Design of SRAM-Based Low-Cost SEU Monitor for Self-Adaptive Multiprocessing System," in Proc. Euromicro Conference on Digital System Design (DSD), 2019.
[15] M. Andjelkovic et al., "A Particle Detectors Based on Pulse Stretching Inverter Chain," in Proc. IEEE International Conference on Electronic Circuits and Systems (ICECS), 2019.
[16] A. Simevski et al., "A Scalable and Configurable Multi-Chip SRAM for Space Applications," in Proc. IEEE International Symposium on Defect and Fault Tolerance in VLSI and Nanotechnology (DFT), 2019.
[17] J. S. Kauppila et al., "A Bias-Dependent Single-Event Compact Model Implemented into BSIM4 and a 90 nm CMOS Process Design Kit," IEEE Trans. on Nuclear Science, 2009.
[18] B. Narasimhan et al., "Characterization of Digital Single Event Transient Pulse Widths in 130 nm and 90 nm CMOS Technologies," IEEE Trans. on Nuclear Science, 2007.